\documentclass[12pt]{article}
\usepackage{amsmath}
\usepackage{graphicx}
\usepackage{natbib}
\usepackage{url}

\graphicspath{{fig/}}
\usepackage{bm}
\usepackage{color,soul}
\usepackage{pdflscape}
\usepackage{algorithm}
\usepackage{algpseudocode,url}
\usepackage{enumerate}
\usepackage{xcolor}
\RequirePackage{amsthm,amsmath,amsfonts,amssymb}
\usepackage[normalem]{ulem}

\def\L{\mathbf{L}}
\def\T{\mathbf{T}}
\def\l{\mathbf{l}}
\def\z{\mathbf{z}}
\def\x{\mathbf{x}}
\def\q{\mathbf{q}}
\def\g{\mathbf{g}}
\def\s{\mathbf{s}}
\def\Z{\mathbf{Z}}
\def\M{\mathbf{M}}
\def\U{\mathbf{U}}
\def\V{\mathbf{V}}
\def\B{\mathbf{B}}
\def\A{\mathbf{A}}
\def\D{\mathbf{D}}
\def\X{\mathbf{X}}
\def\C{\mathbf{C}}
\def\I{\mathbf{I}}
\def\E{\mathbb{E}}
\def\KL{\mathbb{KL}}

\def\G{\mathcal{G}}
\def\zk{\z_k}

\def\Zhat{\hat{\Z}}

\def\tr{\text{tr}}
\DeclareMathOperator*{\argmax}{arg\,max}
\DeclareMathOperator*{\argmin}{arg\,min}
\newcommand{\overbar}[1]{\mkern 1.5mu\overline{\mkern-1.5mu#1\mkern-1.5mu}\mkern 1.5mu}

\newtheorem{lemma}{Lemma}
\newtheorem{proposition}{Proposition}
\newtheorem{theorem}{Theorem}

\newtheorem{fact}{Fact}
\newtheorem{definition}{Definition}

\newcommand{\blind}{0}

\begin{document}

\def\spacingset#1{\renewcommand{\baselinestretch}
{#1}\small\normalsize} \spacingset{1}


\if0\blind
{
  \title{\bf Empirical Bayes Covariance Decomposition, and a Solution to the Multiple Tuning Problem in Sparse PCA}
  \author{Joonsuk Kang
  \hspace{.2cm}\\
    Department of Statistics, University of Chicago\\
    and \\
    Matthew Stephens \\
    Departments of Statistics and Human Genetics, University of Chicago}
  \maketitle
} \fi

\if1\blind
{
  \bigskip
  \bigskip
  \bigskip
  \begin{center}
    {\LARGE\bf Title}
\end{center}
  \medskip
} \fi

\bigskip
\begin{abstract}
Sparse Principal Components Analysis (PCA) has been proposed as a way to improve both interpretability and reliability of PCA. However, use of sparse PCA in practice is hindered by the difficulty of tuning the multiple hyperparameters that control the sparsity of different PCs (the ``multiple tuning problem", MTP). Here we present a solution to the MTP using Empirical Bayes methods. We first introduce a general formulation for penalized PCA of a data matrix $\mathbf{X}$, which includes some existing sparse PCA methods as special cases. We show that this formulation also leads to a penalized decomposition of the covariance (or Gram) matrix, $\mathbf{X}^T\mathbf{X}$. We introduce empirical Bayes versions of these penalized problems, in which the penalties are determined by prior distributions that are estimated from the data by maximum likelihood rather than cross-validation. The resulting ``Empirical Bayes Covariance Decomposition" provides a principled and efficient solution to the MTP in sparse PCA, and one that can be immediately extended to incorporate other structural assumptions (e.g. non-negative PCA).  We illustrate the effectiveness of this approach on both simulated and real data examples.
\end{abstract}

\vfill

\newpage
\spacingset{1.75}

\section{Introduction}

Principal components analysis (PCA, \citealp{Pearson1901}) is a popular dimension reduction technique for revealing structure in data. However, when applied to large data sets, PCA results are often difficult to interpret. To address this, many authors have considered modifications of PCA that use sparsity to help produce more interpretable results. Early versions of this idea arose in the literature on Factor Analysis, where practitioners applied rotations
to post-process results from PCA, or related techniques, to obtain sparse
solutions; see \cite{Rohe.Zeng2023} for interesting background and discussion. More recently, many authors have introduced ``sparse PCA" (sPCA) methods that directly incorporate notions of sparsity into the inference problem \citep[e.g.][]{dAspremont.Ghaoui.ea2004, Zou.Hastie.ea2006, Witten.Tibshirani.ea2009,Journee.Nesterov.ea2010,Ma2013}. 

There exist several different characterizations of PCA, which are equivalent, but lead to different sparse versions  \citep{Zou.Xue2018,Guerra-Urzola.VanDeun.ea2021}. 
One characterization of PCA \cite[section 3.5]{Jolliffe2002}, and the one we focus on here, is that PCA finds a rank-$K$ approximation of an {$N$-by-$P$} data matrix $\mathbf{X}$, 
\begin{equation} \label{eq:x_approx}
    \X \approx \Z\L^T = \sum_{k=1}^K \z_k \l_k^T
\end{equation}
where $\Z$ is an $N$-by-$K$ orthogonal matrix $\Z^T\Z=\I_K$, $\L$ is a $P$-by-$K$ matrix, and $\z_k,\l_k$ denote their $k$th columns respectively. Informally, the orthogonality of $\Z$ means also that
\begin{equation} \label{eq:xxt_approx}
    \X^T\X \approx \L\Z^T \Z\L^T = \L\L^T = \sum_{k=1}^K \l_k \l_k^T.
\end{equation}
Thus, PCA simultaneously provides an approximate decomposition of the data matrix \eqref{eq:x_approx} and of the covariance matrix \eqref{eq:xxt_approx} into a sum of $K$ parts. Similarly, some sparse versions of PCA  \citep[e.g.][]{Witten.Tibshirani.ea2009,Journee.Nesterov.ea2010} provide sparse versions of these decompositions, in which the matrix $\L$ is sparse. Assuming sparsity of $\L$ can improve accuracy of the estimated $\L$, and help make these decompositions more interpretable.

An important practical issue in sPCA is deciding how sparse to make each vector $\l_k$. A natural idea is to have the sparsity of each $\l_k$ be controlled by a hyperparameter that is tuned by cross-validation (CV). However, simultaneously tuning $K$ hyperparameters by CV becomes computationally impractical for moderate $K$ \citep{Zou.Xue2018, Feng.Simon2020}.
We call this the  ``\emph{Multiple Tuning Problem}'' (MTP).

Here we present a novel sPCA method that  solves the MTP by leveraging the empirical Bayes (EB) framework. Within the EB framework, penalties come from priors, whose hyperparameters are learned from data. This approach, which seamlessly integrates hyperparameter tuning into the fitting algorithm, offers a compelling computationally-practical alternative to tuning hyperparameters by CV. The EB approach is quite flexible, and although we focus here on sparsity, it could also be used to impose constraints such as non-negativity on $\L$. Our approach differs from previous EB approaches to matrix factorization \citep{Wang.Stephens2021,Zhong.Su.ea2022} in providing a sparse covariance decomposition \eqref{eq:xxt_approx} as well as a data matrix decomposition \eqref{eq:x_approx}, and so we call it ``Empirical Bayes Covariance Decomposition" (EBCD).

The remainder of the paper is organized as follows. Section \ref{ebcd_sec:ppca} introduces a simple and general penalized PCA criterion and a corresponding optimization algorithm, and shows how the criterion can also be interpretated as a penalized covariance decomposition problem. Section \ref{ebcd_sec:ebcd} introduces the empirical Bayes formulation, Section  \ref{ebcd_sec:practical} discusses practical considerations and Section \ref{ebcd_sec:results} presents empirical results. Finally, Section \ref{sec:discussion} discusses extensions beyond sparsity.

\section{A Penalized PCA Criterion, and its corresponding Penalized Covariance Decomposition criterion}\label{ebcd_sec:ppca}

\subsection{A Penalized PCA Criterion}

To formalize \eqref{eq:x_approx} above, one characterization of PCA \cite[section 3.5]{Jolliffe2002} is that it finds the best rank-$K$ approximation of $\mathbf{X}$ in the sense that it solves the following optimization problem:
\begin{align} \label{ebcd_eq:PCA_1_problem}
 \min_{\substack{\Z\in\mathcal{S}(N,K),\\ \L\in\mathcal{M}(P,K)}}
 \frac{1}{2}\|\mathbf{X}-\Z\L^T\|_F^2
    \quad\text{ subject to } \L^T\L \text{ is diagonal.}
\end{align}
where 
$\mathcal{M}(N,K)$ denote the set of $N$-by-$K$ real matrices, 
$\mathcal{S}(P,K)=\{\mathbf{M}\in\mathcal{M}(P,K) \, : \, \mathbf{M}^T\mathbf{M}=\mathbf{I}_K\}$ denote the set of $P$-by-$K$ orthonormal matrices, and $\|\bm{A}\|_F$ denotes the Frobenius norm of the matrix $\bm{A}$.
The matrices $\Z$ and $\L$ are sometimes called the component score and component loading matrices respectively.
The diagonal restriction on $\L^T\L$ can be equivalently phrased as assuming $\L$ is orthogonal.

Based on \eqref{ebcd_eq:PCA_1_problem}, we propose the following \emph{penalized PCA criterion}, obtained by replacing the orthogonality restriction on  $\L$ with a penalty term, which might, for example, encourage $\L$ to be sparse:
\begin{align} \label{ebcd_eq:penpca}
    \min_{\substack{\Z\in\mathcal{S}(N,K),\\ \L\in\mathcal{M}(P,K)}}
    h_{P,\bm{\lambda}}(\L,\Z; \X):= \left(\frac{1}{2}\|\mathbf{X}-\Z\L^T\|_F^2 + \sum_{k=1}^K P(\l_k; \lambda_k)\right)
\end{align}
where $\l_k$ is the $k$th column of the matrix $\L$ and $P(\cdot; \lambda)$ is a penalty function with hyperparameter $\lambda$ whose value determines the strength of the penalty. 
The multiple tuning problem (MTP) described in the Introduction refers to the difficulty of appropriately tuning the $K$ hyperparameters $\bm{\lambda}=(\lambda_1,\dots,\lambda_K)$.
In this section we assume $\bm{\lambda}$ to be fixed and known; in the next section we address how it can be automatically tuned using EB methods, thus addressing the MTP.

\subsection{A Penalized Covariance Decomposition Criterion} \label{ebcd_sec:pcd_body}

Solving \eqref{ebcd_eq:penpca} yields an approximate decomposition of the data matrix $\X \approx \Z \L^T$ with $\L$ sparse (assuming that the penalty $P$ is sparsity-inducing). As outlined in the Introduction \eqref{eq:xxt_approx}, because of the orthonormality constraint $\Z \in \mathcal{S}(N,K)$, this also yields an approximate decomposition $\X^T \X \approx \L\L^T$ with $\L$ sparse. This result is formalized in the following theorem (see Appendix \ref{ebcd_sec:pcd} for proof):
\begin{theorem} \label{ebcd_thm:equivalence_main} Let $(\hat{\Z}, \hat{\L})$ denote a solution to \eqref{ebcd_eq:penpca}.  Then $\hat{\L}$ also solves
\begin{align}  \label{ebcd_eq:pencov_main}
    \hat{\L} \in \argmin_{\L\in\mathcal{M}(P,K)} \left(\frac{1}{2}d_*(\mathbf{X}^T\mathbf{X}, \L\L^T)^2+\sum_{k=1}^K P(\l_k; \lambda_k)
    \right)
\end{align}
where $d_*$ denotes the Bures-Wasserstein distance between two symmetric positive semi-definite (PSD) matrices \citep{Bhatia.Jain.ea2019}.
\end{theorem}
Theorem \ref{ebcd_thm:equivalence_main} characterizes $\hat\L\hat\L^T$ as an optimal (penalized) approximation to the Gram matrix $\X^T \X$. If $\X$ has centered columns then the Gram matrix is proportional to the covariance matrix, and so we call \eqref{ebcd_eq:pencov_main} a ``penalized covariance matrix criterion". Simultaneously providing an approximation to both $\X^T\X$ and $\X$ could be seen as a fundamental characteristic of PCA that is not generally shared by other matrix factorization methods. For this reason we view the orthonormality constraint on $\Z$, which is crucial to Theorem \ref{ebcd_thm:equivalence_main}, as a fundamental feature that distinguishes our work here from similar approaches that do not have this constraint \citep[e.g.][]{Wang.Stephens2021,Zhong.Su.ea2022}.
 
We emphasize two main contributions of Theorem \ref{ebcd_thm:equivalence_main}. The first is conceptual: it characterizes the sense in which solving \eqref{ebcd_eq:penpca} finds an $\L$ such that the penalty is small and $\L\L^T \approx \X^T\X$. Although our focus here is on sparsity-inducing penalties, the result applies more generally. For example, Theorem \ref{ebcd_thm:equivalence_main} shows that solving \eqref{ebcd_eq:penpca} with a penalty that disallows negative entries in $\L$ \citep[see][]{li2021topic} is a form of symmetric non-negative matrix factorization \citep{he2011symmetric}. The second contribution is computational. The Theorem shows that the optimal $\hat\L$ depends on $\X$ only through $\X^T\X$. Thus, the $\hat\L$ for an observed data matrix $\X_\text{obs}$ can be found by solving \eqref{ebcd_eq:penpca}
using $\X=\C$ for any $\C$ such that $\C^T\C= \X_\text{obs}^T\X_\text{obs}$. This allows $\hat\L$ to be computed given access only to $\X_\text{obs}^T\X_\text{obs}$, and not $\X_\text{obs}$; it could also have computational benefits if $P<<N$ since $\C$ can be a $P \times P$ matrix, much smaller than the $N \times P$ matrix $\X_\text{obs}$. (Computationally, \eqref{ebcd_eq:pencov_main} is less convenient to deal with than \eqref{ebcd_eq:penpca}, so all our algorithms work by solving the latter.)

\subsection{Uniting Previous Sparse PCA Methods}

Although \eqref{ebcd_eq:penpca} provides a natural formulation of sPCA,  most previous  methods have not been explicitly framed as optimizing this criterion; see \cite{van2011flexible} for an exception. Nonetheless, several previous sPCA methods are either equivalent to, or closely-related to, solving \eqref{ebcd_eq:penpca} with some choice of penalty. 

The sparse principal components (SPC) method of \cite{Witten.Tibshirani.ea2009} (their Algorithm 2) can be interpreted as a greedy algorithm for solving \eqref{ebcd_eq:penpca} with an $L_1$ penalty ($P(\l_k; \lambda_k)= \lambda_k \|\l_k\|_1$).
Specifically, their Algorithm 2 solves a rank-1 ($K=1$) version of the problem to obtain the first sparse PC, and then obtains subsequent components by ``deflation" \citep{Mackey2008}.

Similarly, the generalized power (GPower) method of \cite{Journee.Nesterov.ea2010} is closely connected to solving \eqref{ebcd_eq:penpca} with an Elastic Net (EN) penalty \citep{Zou.Hastie2005}, $P(\l_k; \bm{\lambda}_k)=\lambda_{k,1} \|\l_k\|_1 +\lambda_{k,2} \|\l_k\|_2^2$. 
Specifically, with an EN penalty, the criterion \eqref{ebcd_eq:penpca} can be written as
\begin{equation}
    \max_{\{\mu_1,\dots,\mu_K\}} 
    \Bigg(
    \max_{\substack{\Z: \Z^T\Z=\mathbf{I}_K,\\
    \L: \|\l_k\|=\mu_k}} \left(
    \text{tr}\left(\mathbf{X}^T\Z\L^T\right)
    -\sum_{k=1}^K \lambda_{k,1} \|\l_k\|_1  
    \right)
    -\sum_{k=1}^K \left( \frac{1}{2}+\lambda_{k,2}  \right)\mu_k^2
    \Bigg).
\end{equation}
The GPower criterion coincides with the inner maximization over $\Z$ and $\L$ under the restriction $\Z^T\Z=\mathbf{I}_K, \|\l_k\|=\mu_k$. 
(In GPower the column-wise vector norms $\{\mu_1,\dots,\mu_K\}$ are considered as hyperparameters that must be pre-specified, whereas our formulation suggests an alternative approach where $\lambda_{k,2}$ are pre-specified and the $\mu_k$ are maximized over.)

Finally, the USLPCA method of \cite{adachi2016sparse} is closely related to \eqref{ebcd_eq:penpca} with an $L_0$ penalty ($P(\l_k; \lambda) = \lambda\|\l_k\|_0$) and using the same hyperparameter $\lambda$ for all columns; the main difference is that they use a constraint formulation rather than a penalty.

\subsection{BISPCA, a ``Block" Algorithm for Penalized PCA}

A natural strategy for optimizing the penalized PCA criterion \eqref{ebcd_eq:penpca} is block coordinate descent, alternating between minimizing  over $\Z$ with $\L$ fixed (the ``rotation" step) and over $\L$ with $\Z$ fixed (the ``shrinkage" step).  We call this approach the \emph{Block-Iterative-Shrinkage PCA} (BISPCA) algorithm. We briefly outline the two main steps in this algorithm.

{\bf Optimizing over $\Z$ (Rotation step)}.
Optimizing $h_{P,\bm{\lambda}}(\L,\Z;\X)$ over $\Z$ does not depend on the penalty, and has a well-known solution based on the polar decomposition \citep[e.g.][]{Zou.Hastie.ea2006}:$\Zhat(\L,\X):=\text{Polar.U}(\mathbf{X}\L)$, where \text{Polar.U} is defined as follows.
\begin{definition}[$U$ factor of Polar decomposition]\label{ebcd_def:polar}
    For $\M$ any real-valued matrix, with SVD $\M=\U \D\V^T$, define $\text{Polar.U}(\M):=\U\V^T$. 
\end{definition}

{\bf Optimizing over $\L$ (Shrinkage Step).}

To simplify this step we assume that the penalty term is separable, in that $\sum_k P(\l_k;\lambda_k)$ $= \sum_{p,k} \rho(l_{p,k};\lambda_k)$ for some 1-dimensional penalty function $\rho$.
Optimizing $h_{P,\bm{\lambda}}(\L,\Z;\X)$ over $\L$ then splits into $PK$ independent problems, and
\begin{align}
\hat{l}_{p,k} = \argmin_l \left(
\frac{1}{2}(l-\theta_{p,k})^2+ \rho(l; \lambda_k)
\right),
\end{align}
where $\theta_{p,k} := \mathbf{x}_p^T\zk$.
The solution to this problem,
$S_{\rho}(\theta_{p,k}; \lambda_k)$, depends on the penalty function $\rho(\cdot; \lambda_k)$, and is referred to as the ``proximal operator" of $\rho(\cdot;\lambda_k)$. 
It has a closed-form solution for some widely-used penalties. Common examples include the soft thresholding operator for the $L_1$ penalty and the hard thresholding operator for the $L_0$ penalty \citep{Parikh.Boyd2014}.

\subsection{Connections with Other Algorithms}

Table \ref{ebcd_tab:algo} summarizes the BISPCA algorithm, and compares it with previous sPCA algorithms, highlighting their shrinkage, rotation, and deflation steps, as well as the role of penalty hyperparameters.

With no penalty (or constant penalty) BISPCA is a simple variation on the standard ``orthogonal iteration" method for standard PCA \citep{Wilkinson1965,Golub.VanLoan2013}.
Specifically, in this case the BISPCA updates simplify to 
$\Z\gets \text{Polar.U}(\mathbf{X X^T Z})$, whereas standard orthogonal iteration uses $\Z \gets \text{QR.Q}(\mathbf{XX^T}\Z)$ where $\text{QR.Q}$ denotes the orthogonal $\mathbf{Q}$ factor of the QR decomposition. That is, BISPCA simply uses \text{Polar.U} as an alternative orthogonalization to $\text{QR.Q}$.
Under mild conditions, under either of these iterates, the range of $\Z$ converges to the leading eigenspace of $\mathbf{XX^T}$.

With an $L_1$ penalty, the proximal operator $S_\rho$ is soft thresholding, and BISPCA is closely related to both SPC and GPower algorithms, which also alternate between rotation and soft-thresholding steps. The key difference between BISPCA and SPC is that BISPCA optimizes over all columns of $\L$ jointly, whereas SPC optimizes them sequentially (i.e.~in a greedy way).  
The difference between the BISPCA algorithm (with $L_1$ penalty) and GPower is that GPower adds a normalization step ($\l_k \gets \mu_k \l_k/\|\l_k\|_2$), which requires specifying the column norms $\mu_k$. (Recall that GPower corresponds to an EN penalty, which explains this need to specify additional parameters.)

\begin{table*}
\caption{Sparse PCA Algorithms. $S_\rho(\cdot; \lambda_k)$ denotes the proximal operator of the penalty function $\rho(\cdot; \lambda_k)$, and $S_1$ denotes the soft thresholding operator, which is the proximal operator of the $L_1$ penalty. 
We use $S_\rho(\A; \bm{\lambda})$ to denote the vector whose $k$th element is $S_\rho(\bm{a}_k; \lambda_k)$.
The U factor of the polar decomposition is denoted as $\text{Polar.U}$, and the Q factor of the QR decomposition is denoted as $\text{QR.Q}$. $\Z^\perp$ represents an orthonormal basis that is orthogonal to $\Z$. The function $G$ calculates the estimated prior from the empirical Bayes normal means model, and the function $S$ returns the corresponding posterior mean vector (see Definition \ref{ebcd_def:ebnm}).
For simplicity all methods except EBCD-MM are presented in their form for fixed values of the penalty hyperparameters, without tuning; EBCD-MM is self-tuning because it updates the priors $g_k$.}
\label{ebcd_tab:algo}
\begin{tabular}{*{10}{l}}
\hline
Method & Shrinkage Step & Rotation Step & Deflation Step \\
 
\hline
$\begin{array}{l}
\text{BISPCA} \\
\text{(Section \ref{ebcd_sec:ppca}, this paper)}
\end{array}$
&$
\l_k\gets S_\rho(\mathbf{X}^T\zk; \lambda_k)
$
&
$\Z \gets \text{Polar.U}(\mathbf{X}\L)$
&
NA
\\

&\multicolumn{2}{c}{
[equivalently,
$
\L\gets S_\rho(\mathbf{X}^T\text{Polar.U}(\mathbf{XL}); \bm{\lambda})$]
}
&

\vspace{0.1in}
\\
$\begin{array}{l}
\text{SPC}\\ 
\text{\citep{Witten.Tibshirani.ea2009}}
\end{array}$
&$
\l_k\gets S_1(\mathbf{R}_k^T\zk; \lambda_k)
$
&$
\begin{cases}
\bm{\theta}_k \gets \frac{\Z_{k-1}^{\perp T}\mathbf{R}_k\l_k}{\|\Z_{k-1}^{\perp T}\mathbf{R}_k\l_k\|_2}\\
\zk \gets \Z_{k-1}^\perp \bm{\theta}_k
\end{cases}
$
&
$\mathbf{R}_k=\mathbf{X}-\sum_{k'=1}^{k-1}\Z_{k'}\l_{k'}^T$

\vspace{0.1in}
\\

$\begin{array}{l}
\text{GPower}\\
\text{\citep{Journee.Nesterov.ea2010}}
\end{array}$
&$
\begin{cases}
\l_k\gets S_1(\mathbf{X}^T\zk; \lambda_{k,1})\\
\l_k \gets \mu_k \l_k/\|\l_k\|_2
\end{cases}
$
&
$\Z \gets \text{Polar.U}(\mathbf{X}\L)$
&
NA

\vspace{0.1in}
\\

$\begin{array}{l}
\text{ITSPCA}\\
\text{\citep{Ma2013}}
\end{array}$
&\multicolumn{2}{c}{
$\begin{array}{l}
\L\gets \text{QR.Q}(S_\rho(\mathbf{X}^T\mathbf{X}\L; \bm{\lambda}))
\end{array}$
}
&
NA

\vspace{0.1in}
\\

$\begin{array}{l}
\text{EBCD-MM}\\
\text{(Section \ref{ebcd_sec:ebcd}, this paper)}
\end{array}$
& $
\begin{cases}
g_k \gets G(\mathbf{X}^T\zk, 1/\tau, \mathcal{G})\\
\bar{\l}_k \gets S(\mathbf{X}^T\zk, 1/\tau, g_k)\\
\end{cases}$
&
$\Z\gets \text{Polar.U}(\mathbf{X}\overbar{\L})$
&
NA \\

\hline
\end{tabular}
\end{table*}

Finally, we contrast BISPCA with the iterative thresholding sparse PCA (ITSPCA) algorithm of \cite{Ma2013}. Whereas ITSPCA iterates
$\L\gets \text{QR.Q}(S_\rho(\mathbf{X}^T\mathbf{X}\L; \bm{\lambda}))$, BISPCA iterates
$\L\gets S_\rho(\mathbf{X}^T\text{Polar.U}(\mathbf{XL}); \bm{\lambda})$
where $S_\rho(\mathbf{M}; \bm{\lambda})$ denotes applying the proximal operator to each column of the matrix $\mathbf{M}$, that is, $S_\rho(\mathbf{M}; \bm{\lambda})=[S_\rho(\mathbf{m}_1; \lambda_1),$ $\dots,$ $S_\rho(\mathbf{m}_K; \lambda_K)]$. Written this way, the updates appear similar, but with a different order of the shrinkage and orthogonalization steps, and with different orthogonalization approaches (QR.Q vs Polar.U). 
A conceptual advantage of BISPCA is that it is designed to optimize an explicit objective function \eqref{ebcd_eq:penpca}; in contrast ITSPCA is simply an algorithmic modification of orthogonal iteration, and it is unclear what it optimizes or whether it is guaranteed to converge.
Furthermore, because ITSPCA enforces orthogonality after shrinkage, sparsity of the final $\mathbf{L}$ is not guaranteed.

\section{An Empirical Bayes Solution to the MTP}\label{ebcd_sec:ebcd}

An important problem remains: choosing the penalty function and tuning its hyperparameters, particularly when different components require different sparsity levels. Here we propose an Empirical Bayes solution to these problems, in which the penalty is determined by a prior distribution, and the ``tuning" takes place by estimating the prior distribution from the data. This is accomplished by a simple modification of the iterative
BISPCA algorithm.

\subsection{The EBCD Model}

Motivated by the criterion \eqref{ebcd_eq:penpca} we consider the following empirical Bayes (EB) model:
\begin{align}\label{ebcd_eq:model_start}
    \mathbf{X}&=\Z\L^T+\mathbf{E}\\ 
    l_{p,k} &\sim^\text{indep} g_k\in\mathcal{G}\\ 
    e_{n,p} &\sim^\text{iid} N(\cdot; 0, 1/\tau) \label{ebcd_eq:model_end}
\end{align}
where $\Z\in\mathcal{S}(N,K)$, and $\L$ is independent of $\mathbf{E}$. 
We refer to this as an EB model because the column-wise prior distributions $\g:=\{g_k\}_{k=1}^K$ are to be estimated from the data (subject to the constraint that they come from some prespecified prior family $\mathcal{G}$, which may be parametric or nonparametric). 
The model is
closely related to the EBMF model of \cite{Wang.Stephens2021}, and the EB-PCA model of \cite{Zhong.Su.ea2022}, but with the key difference that our model replaces a prior on $\Z$ with an orthonormality constraint. We will show that fitting this model is equivalent to optimizing a penalized criterion \eqref{ebcd_eq:penpca} (or \eqref{ebcd_eq:pencov_main}) with a penalty whose form is estimated from the data. Consequently, it is also equivalent to optimizing a penalized covariance decomposition criterion \eqref{ebcd_eq:pencov_main}. This latter property distinguishes it from the EBMF model, and so we refer to the model \eqref{ebcd_eq:model_start}-\eqref{ebcd_eq:model_end} as the ``Empirical Bayes Covariance Decomposition" (EBCD) model.

\subsection{Fitting the EBCD Model}

\subsubsection{A unified optimization approach: ELBO maximization}

A standard EB approach to fitting  \eqref{ebcd_eq:model_start}-\eqref{ebcd_eq:model_end} would usually be phrased as a two-step procedure: i) estimate ($\hat{\g}, \hat{\Z},\hat{\tau}$) by maximizing marginal log-likelihood
\begin{align}\label{ebcd_eq:EB_step1}
    (\hat{\g}, \hat{\Z},\hat{\tau}) := \argmax_{\g, \Z, \tau} \log \int p(\mathbf{X}|\Z,\L,\tau)p(\L|\g)d\L
\end{align}
and ii) compute the conditional posterior for $\L$,
\begin{align} \label{ebcd_eq:EB_step2}
    \mathbf{\hat{\q}}(\L):=p(\L | \hat{\g}, \hat{\Z},\hat{\tau}, \mathbf{X})
    \propto \hat{\g}(\L) p(\mathbf{X}|\hat{\Z}, \L, \hat{\tau}).
\end{align}
One might typically report the mean of $\hat{\q}$, 
$\hat{\L}:=\E_{\hat{\q}}(\L)$
as a point estimate for $\L$. 

The two-step procedure \eqref{ebcd_eq:EB_step1}-\eqref{ebcd_eq:EB_step2} can be usefully rephrased as solving a single optimization problem (e.g. see Appendix B.1.1 in \cite{Wang.Sarkar.ea2020}):
\begin{equation}
    (\hat{\g}, \hat{\Z},\hat{\tau},\hat\q) = \argmax_{\g \in \G, \Z, \tau, \q} F(\g, \Z, \tau, \q) \label{ebcd_eq:veb}
\end{equation}
where
\begin{equation} \label{ebcd_eq:elbo}
    F(\g, \Z, \tau, \q) := 
    \E_\q \log p(\X|\Z,\L,\tau) 
    - \KL(\mathbf{q}||\g).
\end{equation}
Here, $\q$ can be any distribution on $\L$, $\E_\q$ denotes expectation over $\L$ having distribution $\q$, and $\KL(\mathbf{q}||\g)=
\E_{\mathbf{q}}[\log \frac{\q(\L)}{\g(\L)}]$ 
denotes the KL divergence from $\g$ to $\q$.
The function $F$ is often referred to as the ``evidence lower bound" (ELBO).  (While the ELBO is often used in the context of variational approximations, note that here we do not impose any constraint on $\q$, so optimizing $F$ is exactly equivalent to the two-step EB procedure \eqref{ebcd_eq:EB_step1}-\eqref{ebcd_eq:EB_step2}.)

\subsubsection{Preliminary: EBNM problems}

Similarly to \cite{Wang.Stephens2021}, optimizing $F$ over $\g,\q$ ends up requiring the solution to a simpler EB problem known as the ``empirical Bayes normal means" problem. That is, one needs a function, $\text{EBNM}$, defined as follows.
\begin{definition} \label{ebcd_def:ebnm}
Let $\text{EBNM}(\x,s^2,\G)$ denote a function that returns the EB solution to the following normal means model:
    \begin{align} \label{ebcd_eq:EBNM_model}
    x_p|\eta_p, s^2 &\sim^\text{indep} N(x_p; \eta_p,s^2) \\
    \eta_p &\sim^\text{iid}g \in \G,
\end{align}
for $p=1,\dots,P$ where $\mathcal{G}$ is a pre-specified prior family. More precisely,
\begin{equation}
    \text{EBNM}(\x,s^2,\G) := \argmax_{g \in \G,q} \E_q \log p(x | \eta, s^2) - \KL(q||g)
\end{equation}
where the optimization of $q$ is over all possible distributions on $\mathbf{\eta}=(\eta_1,\dots,\eta_P)$ and $\KL(\mathbf{q}||\g)=
\E_{\mathbf{q}}[\log \frac{\q(\L)}{\g(\L)}]$ 
denotes the KL divergence from $\g$ to $\q$.
\end{definition}
Efficient EBNM solver implementations exist  for a wide range of prior families $\G$; see \cite{willwerscheid2025ebnm}.

\subsubsection{ELBO maximization with EBNM solvers}

With the EBNM function in hand, $F$ can be optimized as in the following Proposition (see Appendix \ref{ebcd_sec:algoproof} for proof). 
\begin{proposition}\label{ebcd_prop:algo}
Maximizing the evidence lower bound $F(\g, \Z, \tau, \q)$ \eqref{ebcd_eq:elbo} subject to $\Z^T\Z=\mathbf{I}_K$ can be achieved by iteratively updating $(\g,\mathbf{q})$, updating $\Z$, and updating $\tau$, as follows:
\begin{align} \label{ebcd_eq:ebnm_step}
&\text{EBNM step: }\quad\text{for each }k\in [K], \quad(g_k,q_k) \gets \text{EBNM}(\mathbf{X}^T\zk, 1/{\tau}, \mathcal{G}),\\
&\text{Rotation step: } \label{ebcd_eq:rotation_step}
\quad {\Z} \gets \text{Polar.U}(\mathbf{X}{\bar{\L}}),\\ \label{ebcd_eq:precision_step} &\text{Precision step: } \quad
{\tau} \gets 
NP/(\|\mathbf{X}-{\Z}{\bar{\L}} \text{}^T\|_F^2+\|{\mathbf{V}}\|_{1,1}).
\end{align}
Here $\bar{\L} = \E_\q(\L)$, $\mathbf{V}$ is the matrix with $v_{p,k}=\text{Var}_{q_k}(l_{p,k})$, and $\|\mathbf{V}\|_{1,1}=\sum_{p=1}^P \sum_{k=1}^K v_{p,k}$.
\end{proposition}
This algorithm is similar to that for EBMF in \cite{Wang.Stephens2021}, but with the important distinction that here, due to the  orthogonality constraint on $\Z$, the EBNM updates for $\l_1,\dots,\l_K$ decouple and can be carried out independently.

To clarify the connection with the penalized PCA algorithms, fix $\tau$ and separate the EBNM step into estimation of the prior and computation of the posterior mean: let 
$G(\x, s^2, \G)$ denote the optimal prior returned by 
$\text{EBNM}(\x,s^2, \G)$, and
$S(\x,s^2,g) := \E(\bm{\eta} | \x, s^2, g)$ denote the corresponding posterior mean.
Then the updates \eqref{ebcd_eq:ebnm_step}-\eqref{ebcd_eq:rotation_step} can be rewritten as
\begin{align} \label{ebcd_eq:est_gk}
g_k &\gets G(\mathbf{X}^T\zk, 1/\tau, \mathcal{G})\\
\bar{\l}_k &\gets S(\mathbf{X}^T\zk, 1/\tau, g_k)\\
\Z &\gets \text{Polar.U}(\mathbf{X}\overbar{\L}).
\end{align}
We call this algorithm EBCD-MM because it can be framed as a ``minorization-maximization" (MM) algorithm to optimize the EBCD criterion, the minorization being given by $F$ in \eqref{ebcd_eq:elbo}. 
Comparing EBCD-MM with BISPCA (Table \ref{ebcd_tab:algo}) we see that in EBCD-MM $S(\mathbf{x},s^2,g)$ plays the same role as the proximal operator in BISPCA. 
For certain classes of prior $\G$, including the point-Laplace prior we use later, $S$ is a shrinkage operator satisfying $|S(\mathbf{x},s^2,g)|\leq |\mathbf{x}|$ elementwise. In EBCD, the form and strength of this shrinkage are determined by the data through the estimated priors $g_k$ and precision $\tau$. Estimating these quantities automatically tunes the penalty parameters, solving the MTP.

\subsection{Connecting EBCD and Penalized Criteria}

The algorithmic similarity between EBCD-MM and BISPCA reflects a deeper connection: for fixed $\g,\tau$, EBCD is a penalized PCA approach, with a penalty that depends on $\g,\tau$. This is formalized in the following Proposition:
\begin{proposition}\label{ebcd_prop:EB_cov_interpretation}
Define 
\begin{equation}
     \tilde{F}(\g,\Z,\tau,\bar{\L}):= \max_{\q:\E_\q(\L)=\bar{\L}} F(\g,\Z,\tau,\q).
\end{equation}
Then
\begin{align}\label{ebcd_eq:EBCD_criterion}
    \tilde{F}(\g,  \Z, \tau, \bar{\L}) 
    =- \left(\frac{1}{2}
\|\mathbf{X}-\Z\overbar{\L}^T\|_F^2 + \sum_{k,p} P_{\tau,g_k}(\bar{l}_{pk})
\right)\tau
\end{align}
where the penalty terms are given by
\begin{align}\label{ebcd_eq:P_tau}
    P_{\tau,g}(\bar{l})
    =
    \frac{N}{2K\tau}\log\frac{2\pi}{\tau}
    + \frac{1}{2}\min_{
                q: \mathbb{E}_{q}[l]=\bar{l}
    } \left(var_{q}(l) + \frac{2}{\tau} \KL(q||g) \right)
    .
\end{align}
\end{proposition}
Although the penalty $P_{\tau,g}$ does not, in general, have a closed form, it does have some convenient properties; for example, its proximal operator $S_\rho$is the posterior mean from a normal means problem, which has a closed form for many choices of prior $g$. See \cite{kim2024flexible} for some other relevant results.

Since EBCD is a penalized PCA method, the ideas from Section \ref{ebcd_sec:pcd_body} apply. Thus EBCD is also a penalized covariance decomposition, and the solution for an observed data matrix $\X_\text{obs}$ can be computed by applying the EBCD-MM algorithm to $\X=\C$ where $\C$ is any matrix such that $\C^T\C = \X_\text{obs}^T\X_\text{obs}$. (The results from Section \ref{ebcd_sec:pcd_body} concern fixed $g,\tau$, but it is straightforward to extend these results to estimating $\g,\tau$).
Note that step \eqref{ebcd_eq:precision_step} requires $N$, the number of rows of $\X_\text{obs}$, which may differ from the number of rows of $\C$.

\subsection{Variations and Extensions}

Appendix \ref{ebcd_sec:ext} describes variations of EBCD that  rescale the orthogonality constraint on $\Z$ (which yields a formulation with a natural population interpretation) and extend the model to allow for column-wise variances.

\section{Practical Issues} \label{ebcd_sec:practical}

\subsection{Initialization} \label{ebcd_sec:init}

Both the penalized PCA criterion and the EBCD criterion are non-convex, so solutions may depend on initialization.
We adopt a simple and effective greedy strategy that sequentially adds columns to $\L$ and $\Z$. Each step is initialized using a rank-one (unpenalized) truncated SVD of the current residual matrix. 
To enforce that the newly added column $\z_k$ is orthogonal to the existing columns of $\Z_{k-1}$, we use the rotation step of SPC \citep{Witten.Tibshirani.ea2009} (see Table \ref{ebcd_tab:algo}) as our `greedy rotation step':
\begin{align}\label{ebcd_eq:greedyrotation}
    \zk \gets &\sqrt{N}\text{greedyrotation}\left(\Z_{k-1}, \mathbf{R}_k, \l_k\right)
    :=\sqrt{N}\frac{\Z_{k-1}^\perp\Z_{k-1}^{\perp T}\mathbf{R}_k\l_k}{\|\Z_{k-1}^{\perp T}\mathbf{R}_k\l_k\|_2}.
\end{align}
After $K$ columns have been added the criterion can be further optimized by applying EBCD-MM, a process referred to as ``backfitting" in \cite{Wang.Stephens2021}. For completeness we give the full procedure in Algorithm \ref{ebcd_algo:ebcdmm}.

\subsection{Choice of $K$} 

As noted in \cite{Wang.Stephens2021}, the EB approach provides a way to automatically select $K$. Provided the prior family $\G$ includes the distribution $\delta_0$, a point mass at 0, then the EBCD criterion may be optimized with some $g_k=\delta_0$, and hence $\bar{\l}_k=0$. Algorithmically, the greedy procedure in Algorithm $\ref{ebcd_algo:ebcdmm}$ can be terminated the first time that $\bar{\l}_0=0$, providing an automatic way to stop adding factors. Alternatively the algorithm can, of course, be run with a user-specified choice of $K$.

\begin{algorithm}[h]
\caption{EBCD-MM (greedy + backfit)}\label{ebcd_algo:ebcdmm}
\begin{algorithmic}
\Require data $\mathbf{X}$ (with $N$ rows and $P$ columns); maximum number of PCs $\textit{Kmax}$; function $\text{svd1}(\mathbf{A})\to (\mathbf{u},d,\mathbf{v})$ that returns the leading singular vectors and singular value;  function ${\tt ebnm}(\mathbf{x},s^2,\mathcal{G})\to (\mathbb{E}_{p^\text{post}}[\bm{\eta}], \text{var}_{p^\text{post}}(\bm{\eta}))$ that solves an empirical Bayes normal means problem and returns posterior mean and variance (see Definition \ref{ebcd_def:ebnm}); function $\text{greedyrotation}(\Z,\mathbf{R},\l)\to \z$ that returns the updated  column $\z_0$ that is orthogonal to the existing columns of $\Z$ (see \eqref{ebcd_eq:greedyrotation}).
\vspace{0.1in}
\spacingset{1}
\State $\Z\gets [\text{ }]$; $\bar{\L} \gets [\text{ }]$; $\tau \gets NP/\|\mathbf{X}\|_F^2$
\Comment{Initialize $(\Z$, $\bar{\L}$, $\tau)$}

\For{$r$ in $1,\dots,\textit{Kmax}$} \Comment{Greedily add components up to $Kmax$}
\State $\mathbf{R}\gets \mathbf{X}-\Z\bar{\L}^T$
\State $(\mathbf{u},d,\mathbf{v}) \gets \text{svd1}(\mathbf{R})$
\State $\bar{\l}_0 \gets d\mathbf{v}/\sqrt{N}$
\State $\z_0 \gets \sqrt{N}\text{greedyrotation}(\Z,\mathbf{R},\bar{\l}_0)$
\Repeat
\State $(\bar{\l}_0,\mathbf{v}_0)\gets {\tt ebnm}(\mathbf{R}^T\z_0/N, 1/N\tau, \mathcal{G}_L)$ 
 \Comment{Shrinkage Step}
\State $\z_0 \gets \sqrt{N}\text{greedyrotation}(\Z,\mathbf{R},\bar{\l}_0)$
\Comment{Greedy Rotation Step}
\State $\tau \gets NP/(\|\mathbf{R}-\z_0\bar{\l}_0^T\|_F^2+N\|\mathbf{v}_0\|_{1})$ \Comment{Precision Step}
\Until{convergence criterion satisfied}
\State $\bar{\L} \gets [\bar{\L}, \bar{\l}_0]$
\State $\Z \gets \sqrt{N}\text{Polar.U}(\mathbf{X}\bar{\L})$
\EndFor

\Repeat \Comment{Backfit}
\For{$k$ in $1,\dots,\textit{Kmax}$} \Comment{Shrinkage Step}
\State $(\bar{\l}_k,\mathbf{v}_k)\gets {\tt ebnm}(\mathbf{X}^T\zk/N, 1/N\tau, \mathcal{G}_L)$ 
\EndFor
\State $\Z\gets \sqrt{N}\text{Polar.U}(\mathbf{X}\bar{\L})$
\Comment{Rotation Step}
\State $\tau \gets NP/(\|\mathbf{X}-\Z\bar{\L}^T\|_F^2+N\|\mathbf{V}\|_{1,1})$ \Comment{Precision Step}
\Until{convergence criterion satisfied}

\State \Return $(\Z$, $\bar{\L}$, $\mathbf{V}$, $\tau)$

\end{algorithmic}
\end{algorithm}

\spacingset{1.75}

\subsection{Choice of Prior}

The posterior mean shrinkage operator $S$ in EBCD depends on the prior $g$, thus on the choice of the prior family $\mathcal{G}$. For {\it sparse} PCA one would choose a sparsity-inducing prior family; one could alternatively use non-negative prior families to induce non-negative PCA, or fully nonparametric prior families (as in \cite{Zhong.Su.ea2022}) for a more flexible regularized PCA, although we do not explore these options further here.

\begin{figure*}[h]
\centering
\makebox{\includegraphics[width=0.8\linewidth]{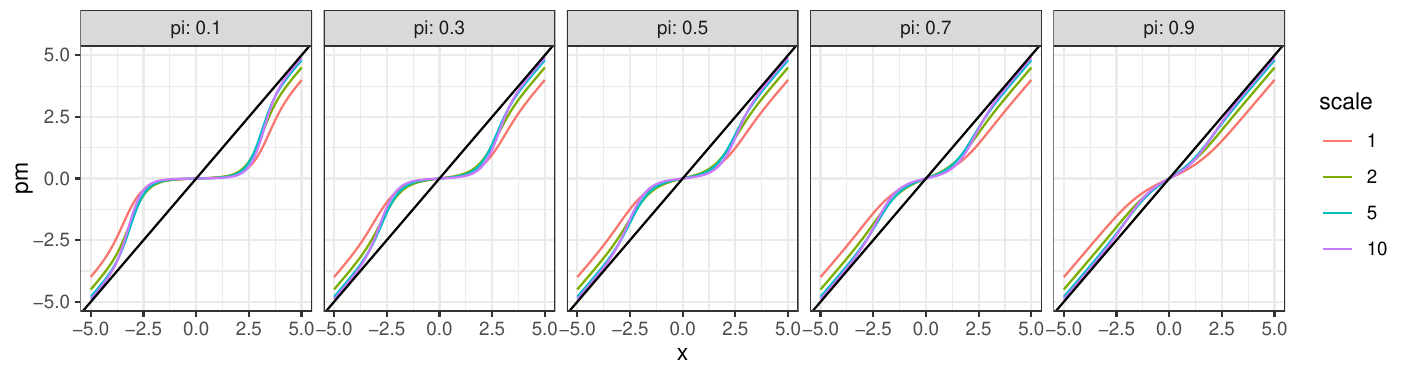}}
\caption{\label{ebcd_fig:pl}Examples of posterior mean shrinkage operator $S(\mathbf{x}, s^2=1, g=g(\cdot; \pi, b))$ induced by Laplace slab priors $g(x; \pi,b)=(1-\pi)\delta_{0}(x)+\pi\text{Laplace}(x; 0, b)$. Note how $\pi$ controls shrinkage near 0 (small $\pi$ yielding more shrinkage), while the scale parameter controls shrinkage further away from 0.}
\end{figure*}

While several choices of sparse family are possible, here we use the ``point Laplace" prior, a spike and slab prior with Laplace slab, as it offers a wide range of shrinkage behaviors while remaining computationally convenient: 
\begin{align}
    \mathcal{G} = \big\{g: &
    g(x)=(1-\pi)\delta_{0}(x)+\pi\text{Laplace}(x; 0, b) \text{ for some } \pi\in [0,1], b>0 \big\}
\end{align}
where $\text{Laplace}(\cdot; \mu,b)$ denotes the Laplace density with location $\mu$ and scale $b$. 
Varying the prior parameters $(\pi,b)$ allows for flexible shrinkage behaviors (Figure \ref{ebcd_fig:pl}). 
We refer to EBCD with this specific prior as EBCD-pl.

\section{Empirical Results}\label{ebcd_sec:results}

\subsection{Simulation}

We compare EBCD-pl with several competing methods, including PCA, $L_1$-penalized PCA (our penalized PCA criterion \eqref{ebcd_eq:penpca} with an $L_1$ penalty), SPC, GPower, and empirical Bayes matrix factorization. 
To avoid the MTP, we implemented a single tuning parameter for $L_1$-penalized PCA and GPower; we applied SPC with a deflation scheme and greedy hyperparameter optimization. 
Following \cite{Journee.Nesterov.ea2010} we fix GPower hyperparameters $(\mu_1,\mu_2)=(1, 0.5)$.

We also compare with EBCD-l (EBCD with column-wise Laplace priors), EBMF-n/l (EBMF with column-wise normal priors on $\mathbf{Z}$ and column-wise Laplace priors on $\mathbf{L}$), EBMF-n/pl (EBMF with column-wise normal priors on $\mathbf{Z}$ and column-wise point-Laplace priors on $\mathbf{L}$), and empirical Bayes PCA (EB-PCA). 
Although we believe that Laplace priors are not sufficiently expressive for sparse PCA, we include EBCD-l and EBMF-n/l to facilitate comparison with $L_1$-penalty-based methods. Similarly, we include EBMF-n/pl to compare with EBCD-pl and assess the impact of orthogonality restriction. 

We consider two simulation settings, each with $\x_1,\dots,\x_{50}$ $\sim N_{500}(\bm{0}, \bm\Sigma)$, where the $500 \times 500$ covariance matrix $\bm\Sigma$ is given by:
\begin{equation}\text{Setting 1.}\quad\quad\nonumber
\bm{\Sigma}=399\mathbf{v}_1\mathbf{v}_1^T+299\mathbf{v}_2\mathbf{v}_2^T+\mathbf{I}_{500}
\end{equation}
where the PCs $\mathbf{v}_1,\mathbf{v}_2$ are given by
    $v_{1,j}=\mathbf{1}_{j\in [1,10]}/\sqrt{10}$ and 
    $v_{2,j}=\mathbf{1}_{j \in [11,20]}/\sqrt{10}$.
 This setting comes from \cite{Shen.Huang2008} and \cite{Journee.Nesterov.ea2010};
\begin{equation}\text{Setting 2.}\quad\quad\nonumber
\bm{\Sigma}=9\mathbf{v}_1\mathbf{v}_1^T+7\mathbf{v}_2\mathbf{v}_2^T+4\mathbf{v}_3\mathbf{v}_3^T+\mathbf{I}_{500}
\end{equation}
where $v_{1,j}=\mathbf{1}_{j\in [1,10]}/\sqrt{10}$, $v_{2,j}=\mathbf{1}_{j \in [11,50]}/\sqrt{40}$, and $v_{3,j}=\mathbf{1}_{j \in [51,150]}/\sqrt{100}$.
This setting illustrates the effect of non-equal sparsity level in the PCs.

For each setting we simulate 50 datasets and measure  performance by three measures: i) the angle between the true PC and its estimate: for each PC $i$, the angle is defined as
$d_i = \angle(\mathbf{v}_i,\hat{\l}_i)/\frac{\pi}{2}$
where $\angle(\cdot, \cdot)$ denotes the angle between two vectors; ii) the difference between the population covariance matrix and the estimated $\frac{1}{N}\hat{\L}\hat{\L}^T$: $d_\text{cov} = \|\bm{\Sigma}-\frac{1}{N}\hat{\L}\hat{\L}^T\|_F;$
iii) the distance with optimal rotation, which measures the proximity of two subspaces: $d_\text{or} =\min_{\mathbf{R}\in\mathcal{O}^{K \times K}}\|\tilde{\L} \mathbf{R}-\mathbf{V}\|_F$
where $\mathcal{O}^{K \times K}$ is the set of $K$-by-$K$ orthonormal matrices, $\mathbf{V}$ is $[\mathbf{v}_1,\mathbf{v}_2]$ in Simulation 1 and $[\mathbf{v}_1,\mathbf{v}_2, \mathbf{v}_3]$ in Simulation 2, and $\tilde{\L}$ is an orthonormal basis of the subspace spanned by estimated loading $\hat{\L}$. 

The run-time for EBCD-pl was comparable to that of other sPCA methods. (In Simulation 1, average run-times for each dataset were:  EBCD-pl 2.40s, EBCD-l 0.56s, EBMF-n/pl 0.19s, EBMF-n/l 0.20s, SPC 1.47s, GPower 0.11s, $L_1$-penalized PCA 0.43s, EB-PCA 0.28s, and PCA 0.01s.  For Simulation 2, average run-times were EBCD-pl 2.33s, EBCD-l 2.45s, EBMF-n/pl 0.25s, EBMF-n/l 0.19s, SPC 2.44s, $L_1$-penalized PCA 1.51s, EB-PCA 0.64s, and PCA 0.01s.)
Note that $L_1$-penalized PCA and GPower with equality restriction were optimized over a one-dimensional hyperparameter grid, not over a two-dimensional or three-dimensional grid, which could increase the run-time substantially. Runtimes can vary across software environments, so comparisons should be interpreted qualitatively.

\begin{figure}[h]
\centering
\makebox{\includegraphics[width=\linewidth]{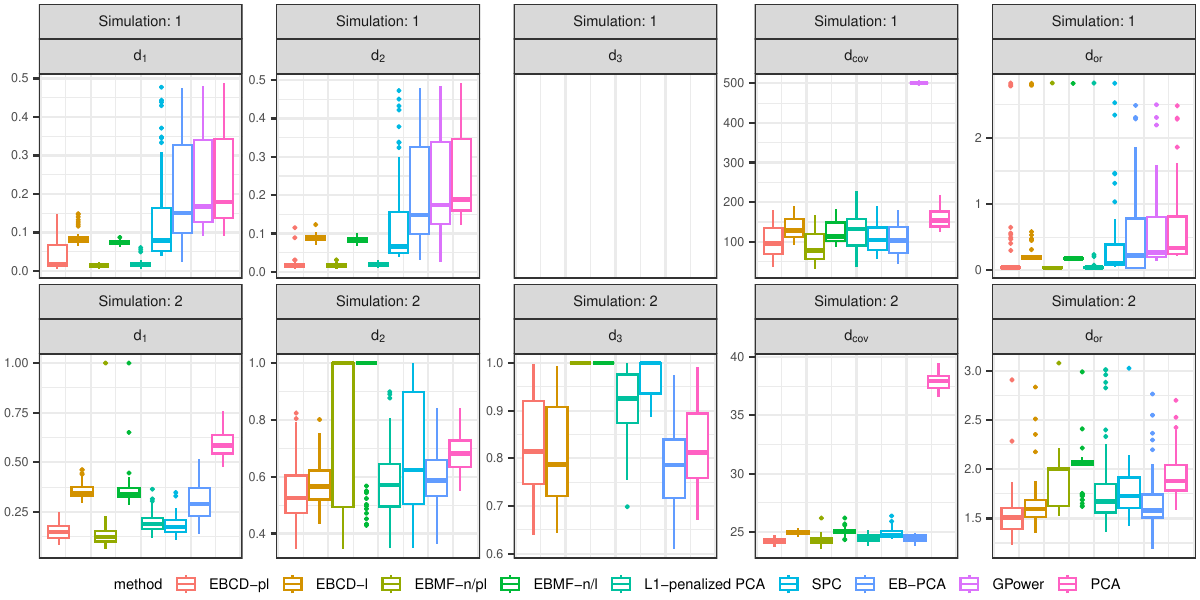}}
\caption{\label{ebcd_fig:sim}Simulation results comparing the performance of different methods in terms of three measures: angle between true and estimated principal components (PCs), difference between population covariance matrix and estimated covariance matrix, and distance with optimal rotation.}
\end{figure}

Figure \ref{ebcd_fig:sim} summarizes the results.
EBCD-pl outperforms other methods, with $L_1$-penalized PCA ranking second. The benefits of EBCD-pl over $L_1$-penalized PCA are most pronounced in Simulation 2, where the true PCs have different sparsity levels. However, even in this setting, the performance of $L_1$-penalized PCA is impressive despite the equality restriction on the penalty.
The superiority of $L_1$-penalized PCA over SPC is presumably due in part to 
its use of a joint optimization scheme in place of a greedy/sequential approach. Its superiority compared with the GPower method
may reflect difficulty in selecting the
hyperparameter $\bm{\mu}$ in GPower. Indeed, we excluded GPower results in Simulation 2 as we found it hard to specify this parameter.

The performance differences between EBCD-pl and EBCD-l highlight the importance of selecting a flexible prior family. 
While EBCD-l and $L_1$-penalized PCA perform similarly in Simulation 1, EBCD-l tends to outperform in Simulation 2, underscoring the increasing restrictiveness of the equality constraint. 
In Simulation 1, EBMF-n/pl and EBMF-n/l perform similarly to EBCD-pl and EBCD-l, respectively. However, in Simulation 2 their performance declines due to the omission of one, two, or even all three factors during estimation. This result illustrates the impact of the orthogonality assumption in EBCD.

\subsection{Stock Market Data} \label{sec:stock_market}

To illustrate our method’s effectiveness in producing interpretable results, we applied EBCD-pl to S\&P500 sector returns from January 1, 2020, to November 29, 2022, covering the COVID pandemic. The data consist of daily log returns for $N=734$ trading days across $P=11$ Global Industry Classification Standard (GICS) sectors. The time period and GICS sectors match those analyzed in an article in The \cite{Economist_2022_sectors}.

\begin{figure}[tb]
\centering
\makebox{\includegraphics[width=\linewidth]{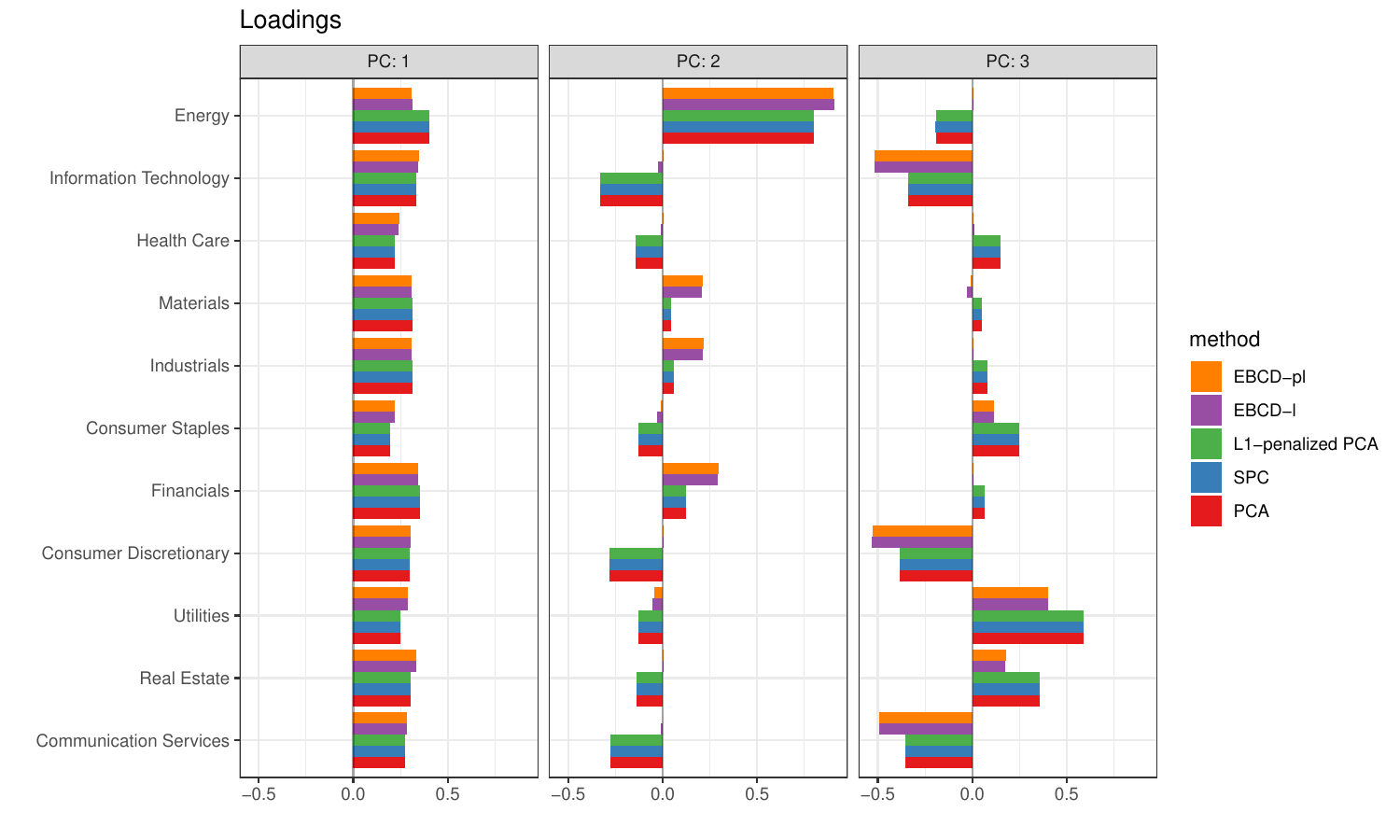}}
\caption{\label{ebcd_fig:sectors_loadings} Comparison of PCA loadings with posterior mean loadings from EBCD-pl (after post-processing to have unit norm).}
\end{figure}

We applied EBCD-pl, EBCD-l, $L_1$-penalized PCA, SPC and classical PCA to these data. The first three classical PCs explain 90.54\% of total variance, with a sharp drop-off in signal after this point (the first five PCs explain 72.49\%, 11.91\%, 6.14\%, 2.40\%, and 1.66\%) and so we focus comparisons on the first three PCs. The $L_1$-penalized PCA and SPC results are almost identical to the PCA result. In contrast the three PCs estimated by EBCD-pl differ from classical PCA, both in their PVEs  (66.99\%, 16.35\%, and 7.13\%) and in the qualitative features of their loadings after the first PC (Figure \ref{ebcd_fig:sectors_loadings}).

The results of EBCD-pl and EBCD-l are similar for these data, but EBCD-pl produces slightly sparser results, as expected given the differences in the expressiveness of the two prior distributions. For example, the loadings for information technology and consumer staples in PC2, and the loadings for health care and materials in PC3, illustrate this difference.
Comparing EBCD-l with $L_1$-penalized PCA, the equality restriction appears too restrictive in this setting, as it fails to account for the varying sparsity levels across the principal components.

We attribute the difference in behavior between EBCD-pl (and EBCD-l) vs SPC as primarily due to the fact that SPC uses a greedy/sequential optimization approach, whereas EBCD-pl performs joint optimization. When signal is strong,  sequentially estimated sparse PCs may not deviate much from classical PCs. In contrast, the joint optimization in the EBCD-pl algorithm
(backfitting stage in Algorithm \ref{ebcd_algo:ebcdmm}) allows
EBCD-pl to move some of the explanatory power of the first PC to other PCs in order to increase sparsity.
Interestingly this is done at almost no expense of total PVE explained by the first three PCs: cumulatively, the three EBCD-pl PCs explain 90.47\% of the variation, very similar to the 90.54\% of classical PCA. 
This highlights a benefit of joint optimization methods for sparse PCA, compared with the widely-used sequential schemes.

The first PC (both classical and EBCD-pl) loads roughly equally on all sectors, and so captures the tendency of sectors to move together as the market varies. To describe the loadings on the second and the third EBCD-pl PCs, we group the sectors into four groups: energy, materials, industrials, and financials (EMIF); consumer staples, utilities, real estate (SUR); information technology, consumer discretionary, and communication services (TDC); and health care. 
The second EBCD-pl PC captures the EMIF sector, and the third EBCD-pl PC captures the contrast between SUR and TDC.

\begin{figure*}[tb]
\centering
\makebox{\includegraphics[width=0.7\linewidth]{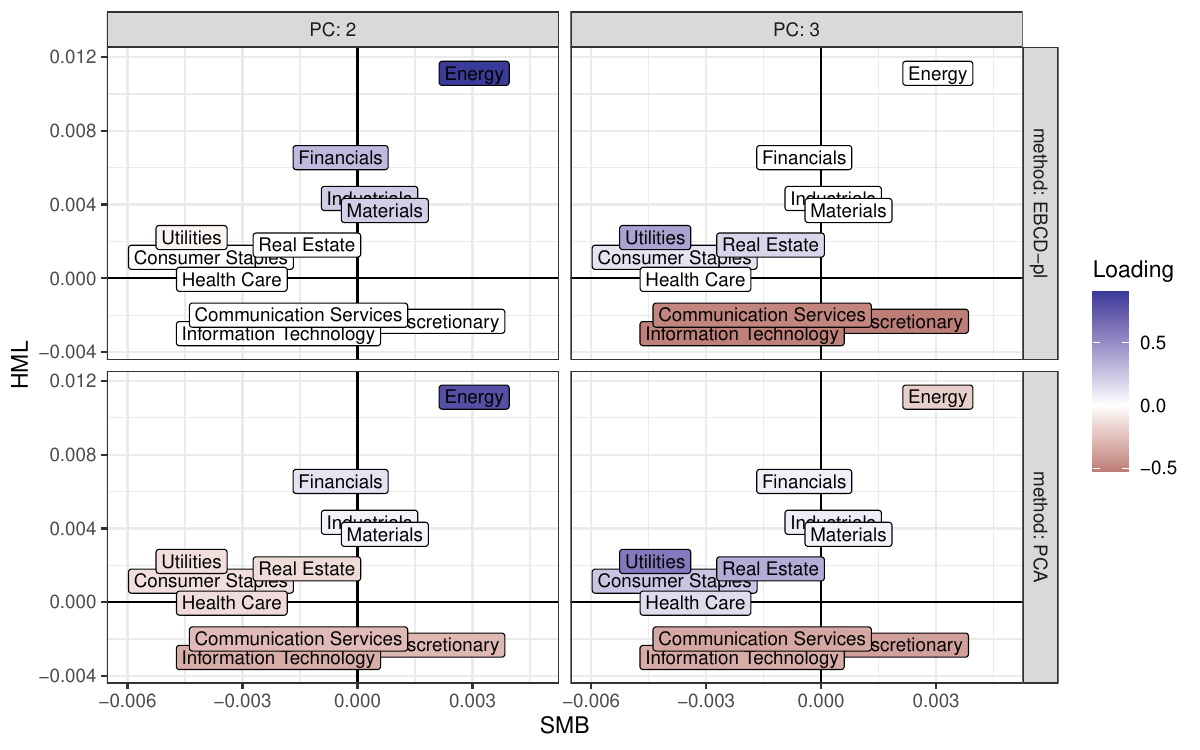}}
\caption{\label{ebcd_fig:sectors_plane}Sectors projected on the SMB-HML plane. Each sector is positioned according to its loadings on the Fama-French SMB and HML factors, and is colored based on its loadings on the second and third principal components (PCs) from the EBCD-pl method (or PCA).}
\end{figure*}

These EBCD-pl results can be interpreted in the context of the 
Fama-French three-factor model \citep{Fama.French1993}, which is the standard model in finance that explains variation in stock prices by three factors:
the market factor (roughly, overall average performance of all stocks), the size factor (SMB, for small minus big, contrasting stocks with small vs big market capitalization), and the growth/value factor (HML, for high minus low, contrasting high value stocks, which have high book-to-market value ratio, with growth stocks which have low book-to-market ratio). The first EBCD-pl PC captures the
market factor, 
whereas the second and third PCs partition the sectors into three groups: the TDC group contains the growth sectors; the EMIF group contains the strong value sectors with smaller sizes and the SUR group contains the moderately value sectors with larger sizes. This is illustrated graphically in Figure \ref{ebcd_fig:sectors_plane}, which shows each sector in the Fama-French SMB-HML plane (data from the Data Library maintained by Kenneth R. French), colored according to loading on 
the second and third PCs. The colorings for EBCD-pl PCs clearly capture contiguous regions of the plane. In contrast the classic PCs do not align so closely with the Fama-French factors; in particular the third PC groups the energy sector with TDC, which do not fall together in the SMB-HML plane.

\section{Discussion} \label{sec:discussion}

We introduced a simple penalized PCA criterion, \eqref{ebcd_eq:penpca} that unites some existing sparse PCA methods (SPC and GPower). We showed that this criterion has the property of simultaneously providing a decomposition of both the data matrix and the covariance, or Gram, matrix. To address the challenge of tuning multiple hyperparameters, we proposed an empirical Bayes approach that integrates hyperparameter tuning directly within the algorithm. The result is an empirical Bayes approach to covariance decomposition (EBCD), which we found in simulations can outperform existing methods for sparse PCA.

While we have focused here on sparsity, our EBCD approach is quite general, and other structures can be easily incorporated simply by changing the prior family used. For example, 
replacing the point-Laplace prior family we used here with a point-Exponential prior family immediately leads to a new EB method for sparse, non-negative PCA \citep{zass2006nonnegative} and, simultaneously, a version of semi-nonnegative matrix factorization \citep{Ding.Li.ea2010}. The non-negative constraint may provide more interpretable covariance decompositions in many applications; see \cite{li2021topic} for interesting recent work in this direction.
Another interesting possibility to improve interpretation is to use binary or near-binary priors, which would lead to empirical Bayes versions of additive clustering \citep{Shepard.Arabie1979}; see also \cite{kueng2021binary,sorensen2022overlapping, kolomvakis2023robust,Liu.Carbonetto.ea2023}. 
Similarly, one could obtain an EB version of ``functional PCA" \citep{ramsay2005principal} by replacing the sparse prior with a ``spatial" prior that encourages $|\eta_i-\eta_{i+1}|$ (in Definition \ref{ebcd_def:ebnm}) to be typically small. EBNM solvers for a range of priors are implemented in the EBNM package \citep{willwerscheid2025ebnm}, and an EBNM solver for a spatial prior is implemented using wavelet methods in \cite{xing2021flexible}, and any of these could be immediately plugged into Algorithm \ref{ebcd_algo:ebcdmm}. It is, however, possible that some prior families may require careful attention to initialization to yield good performance.

\bibliographystyle{chicago}
\bibliography{cited_refs}

\newpage
\appendix
\pagenumbering{arabic}
\setcounter{page}{1}
\renewcommand{\thepage}{A-\arabic{page}}

\section{Data and Resources}

EBCD-pl is implemented in the R package \texttt{ebcd} that is available from \url{https://github.com/joonsukkang/ebcd}. Source code for the empirical results is available from \url{https://github.com/joonsukkang/ebcd-paper}.

In our comparisons we used the R package \texttt{PMA} implementing SPC, the MATLAB implementation  
\url{http://www.montefiore.ulg.ac.be/~journee/GPower.zip} for GPower, the R package \texttt{flashier} for EBMF, and the Python implementation of EB-PCA available on \url{https://github.com/TraceyZhong/EBPCA}.

The sector-level daily returns data was provided by Refinitiv via Wharton Research Data Services. Data will be shared on request to the corresponding author with permission of Refinitiv. The Fama-French 3 Factor Returns data is available in the Data Library maintained by Kenneth R. French: \url{https://mba.tuck.dartmouth.edu/pages/faculty/ken.french/data_library.html}.

\section{A Penalized Covariance Decomposition Criterion}\label{ebcd_sec:pcd}

Here we provide more details on the connection between penalized PCA and penalized covariance decomposition. We begin by proving Theorem \ref{ebcd_thm:equivalence_main}.

\subsection{Proof of Theorem \ref{ebcd_thm:equivalence_main}}\label{ebcd_sec:thmproof}

We begin by establishing the following two Lemmas.

\begin{lemma}\label{ebcd_lem:nucnorm}
For any real-valued matrices $\mathbf{A}\in\mathcal{M}(N_1,N_2)$ and
$\mathbf{B}\in\mathcal{M}(N_2,N_3)$, \\
(a) $\|\mathbf{A}\|_*=\text{tr}(\mathbf{A}^T\text{Polar.U}(\mathbf{A}))$.\\
(b) $\|\mathbf{A}\|_*=\text{tr}(\sqrt{\mathbf{A}\mathbf{A}^T})$.\\
(c) $\|\mathbf{AB}\|_*=\|\sqrt{\mathbf{A}^T\mathbf{A}}\sqrt{\mathbf{B}\mathbf{B}^T}\|_*$.
\end{lemma}

\begin{proof}
Let $\mathbf{U}_A\mathbf{D}_A\mathbf{V}_A^T$ and $\mathbf{U}_B\mathbf{D}_B\mathbf{V}_B^T$ denote the SVDs of $\mathbf{A}$ and $\mathbf{B}$ respectively.
(a) From Definition \ref{ebcd_def:polar}, $\text{Polar.U}(\mathbf{A})=\mathbf{U}_A\mathbf{V}_A^T$; 
$\text{tr}(\mathbf{A}^T \text{Polar.U}(\mathbf{A}))$\\$=\text{tr}(\mathbf{V}_A\mathbf{D}_A\mathbf{U}_A^T\mathbf{U}_A\mathbf{V}_A^T)=\text{tr}(\mathbf{D}_A)=\|\mathbf{A}\|_*$.\\
(b) $\text{tr}(\sqrt{\mathbf{A}\mathbf{A}^T})
=\text{tr}(\sqrt{\mathbf{U}_A\mathbf{D}_A\mathbf{V}_A^T\mathbf{V}_A\mathbf{D}_A\mathbf{U}_A^T})
=\text{tr}(\mathbf{U}_A\mathbf{D}_A\mathbf{U}_A^T)$\\$
=\text{tr}(\mathbf{D}_A)=\|\mathbf{A}\|_*
$.
(c) Since the nuclear norm is unitarily invariant, we have $\|\mathbf{A}\mathbf{B}\|_*
=\|\mathbf{U}_A\mathbf{D}_A\mathbf{V}_A^T\mathbf{U}_B\mathbf{D}_B\mathbf{V}_B^T\|_*
=\|\mathbf{D}_A\mathbf{V}_A^T\mathbf{U}_B\mathbf{D}_B\|_*=\|\mathbf{V}_A\mathbf{D}_A\mathbf{V}_A^T\mathbf{U}_B\mathbf{D}_B\mathbf{U}_B^T\|_*$\\$=\|\sqrt{\mathbf{A}^T\mathbf{A}}\sqrt{\mathbf{B}\mathbf{B}^T}\|_*$.
\end{proof}

\begin{lemma} \label{ebcd_lem:wbdist1}
Let $\mathbf{X}\in\mathcal{M}(N,P)$ and $K$ be a positive integer with $K \leq \min(N,P)$. 
Then
\begin{align}
   \min_{\substack{\Z\in\mathcal{S}(N,K)}}
    \|\mathbf{X}-\Z\L^T\|_F^2     = d_*(\mathbf{X}^T\mathbf{X}, \L\L^T)^2
\end{align}
where, denoting the unique positive semidefinite square root of positive semidefinite matrices $\A$ and $\B$ by $\sqrt{\A}$ and $\sqrt{\B}$,
\begin{equation}\label{ebcd_eq:BW}
d_*(\mathbf{A}, \mathbf{B}):=\left(\text{tr}(\mathbf{A})-2\text{tr}(\sqrt{\sqrt{\mathbf{A}}\mathbf{B}\sqrt{\mathbf{A}}})+\text{tr}(\mathbf{B})\right)^{1/2}
\end{equation}
is the Bures-Wasserstein distance between matrices $\A$ and $\B$, which is a metric on the space of positive semi-definite matrices \citep{Bhatia.Jain.ea2019}. 
\end{lemma}

\begin{proof}

From the well-studied solution that $\hat{\Z}(\L,\X)=\text{Polar.U}(\X\L)$, we have $h(\X,\L)=\text{\tr}(\X^T\X)+\text{tr}(\L\L^T)-2\text{tr}(\L^T\X^T\text{Polar.U}(\X\L))$.
The last term is equal to $-2\|\X\L\|_*$ from Lemma \ref{ebcd_lem:nucnorm}(a), 
to \\
$-2\|\sqrt{\X^T\X}\sqrt{\L\L^T}\|_*$ from Lemma \ref{ebcd_lem:nucnorm}(c),
and to \\
$-2\text{tr}(\sqrt{\sqrt{\X^T\X}\L\L^T\sqrt{\X^T\X}})$
from Lemma \ref{ebcd_lem:nucnorm}(b).
Therefore, $h(\X,\L)=\text{\tr}(\X^T\X)+\text{tr}(\L\L^T)$\\$-2\text{tr}(\sqrt{\sqrt{\X^T\X}\L\L^T\sqrt{\X^T\X}})=d_*(\mathbf{X}^T\mathbf{X}, \L\L^T)^2$.
\end{proof}
Now, we are ready to prove the Theorem \ref{ebcd_thm:equivalence_main}, which follows as a direct corollary of Lemma \ref{ebcd_lem:wbdist1}.

\begin{proof}
Let $(\hat{\Z}, \hat{\L})$ denote a solution to the penalized PCA criterion \eqref{ebcd_eq:penpca}.
That is,
\begin{align}\label{ebcd_eq:proofmain}
    &\frac{1}{2}\|\mathbf{X}-\hat{\Z}\hat{\L}^T\|_F^2+\sum_{k=1}^K P(\hat{\l}_k; \lambda_k) \\\nonumber
    &= \min_{\substack{\Z\in\mathcal{S}(N,K),\\ \L\in\mathcal{M}(P,K)}}
    \left(\frac{1}{2}\|\mathbf{X}-\Z\L^T\|_F^2+\sum_{k=1}^K P(\l_k; \lambda_k) \right).
\end{align}
Since $\hat{\Z}$ is the minimizer of $\|\mathbf{X}-\Z\hat{\L}^T\|_F^2$ by construction, the LHS of \eqref{ebcd_eq:proofmain} is equal to 
\begin{align} \label{ebcd_eq:lhs}
    &\frac{1}{2}\min_{\Z\in\mathcal{S}(N,K)}\|\mathbf{X}-\Z\hat{\L}^T\|_F^2+\sum_{k=1}^K P(\hat{\l}_k; \lambda_k) \\\nonumber
    &=\frac{1}{2}d_*(\X^T\X, \hat{\L}\hat{\L}^T)^2+\sum_{k=1}^K P(\hat{\l}_k; \lambda_k)
\end{align}
by Lemma \ref{ebcd_lem:wbdist1}.
Similarly, by Lemma \ref{ebcd_lem:wbdist1}, the RHS of \eqref{ebcd_eq:proofmain} is equal to
\begin{align} \nonumber
    &\min_{\L\in\mathcal{M}(P,K)}
    \left(\frac{1}{2}\min_{\Z\in\mathcal{S}(N,K)}\|\mathbf{X}-\Z\L^T\|_F^2+\sum_{k=1}^K P(\l_k; \lambda_k) \right)\\ \label{ebcd_eq:rhs}
    &= \min_{\L\in\mathcal{M}(P,K)}
    \left(\frac{1}{2}d_*(\X^T\X, \L\L^T)^2+\sum_{k=1}^K P(\l_k; \lambda_k) \right).
\end{align}
Equating the right-hand-sides of \eqref{ebcd_eq:lhs} and \eqref{ebcd_eq:rhs} shows that $\hat{\L}$ is a solution to the penalized covariance decomposition criterion \eqref{ebcd_eq:pencov_main}.
\end{proof}

\subsection{Efficient Computation} \label{ebcd_sec:suff}

In this section, we formalize the idea that one can estimate $\L$ using only the $P \times P$ matrix $\mathbf{X}^T\mathbf{X}$, without using the $N \times P$ matrix $\mathbf{X}$. This could be useful in settings where $\mathbf{X}$ is unavailable (eg in some genetic studies the genotype matrix $\mathbf{X}$ is unavailable for privacy reasons); it may also be computationally convenient in settings where $P\ll N$.

In brief, as outlined in the main text the idea is to compute the solution $\hat{\L}$ by applying the BISPCA algorithm to a \emph{compact version of the data matrix} $\C\in\mathcal{M}(P,P)$ that satisfies $\C^T\C=\mathbf{X}^T\mathbf{X}$.  If the original matrix $\mathbf{X}$ is available then one can use it to compute the corresponding $\hat{\Z}$.
The following theorem formalizes this approach.

\begin{theorem}\label{ebcd_thm:reformulation}
Suppose that a data matrix $\mathbf{X}\in\mathcal{M}(N,P)$ has the thin singular value decomposition  $\mathbf{U}_X\mathbf{D}_X\mathbf{V}_X^T$ with $P<N$ and $K$ is a positive integer with $K \leq P$. Let $\C\in\mathcal{M}(P,P)$ satisfy $\C^T\C=\mathbf{X}^T\mathbf{X}$ (eg, one such matrix is $\C=\mathbf{V}_X\mathbf{D}_X\mathbf{V}_X^T$).
The following four problems are equivalent:
\begin{align}\nonumber
    (a) \quad&\hat{\L},\hat{\Z} \in \argmin_{\substack{\Z\in\mathcal{S}(N,K),\\ \L\in\mathcal{M}(P,K)}}
    \left(\frac{1}{2}\|\mathbf{X}-\Z\L^T\|_F^2
    + \sum_{k=1}^K P(\l_k; \lambda_k)\right)
    \\\nonumber
    (b)\quad  &\hat{\L} \in \argmin_{\L\in\mathcal{M}(P,K)} 
    \left(\frac{1}{2}d_*(\mathbf{X}^T\mathbf{X}, \L\L^T)^2
    + \sum_{k=1}^K P(\l_k; \lambda_k)\right)
    \\\nonumber &\text{and set }\hat{\Z}=\text{Polar.U}(\mathbf{X}\hat{\L})\\\nonumber
    (c)\quad &\hat{\L} \in \argmin_{\L\in\mathcal{M}(P,K)} 
    \left(\frac{1}{2} d_*(\C^T\C, \L\L^T)^2
    + \sum_{k=1}^K P(\l_k; \lambda_k)\right)
    \\\nonumber&\text{and set }\hat{\tilde{\Z}}=\text{Polar.U}(\C\hat{\L}), \hat{\Z}=\mathbf{U}_X\mathbf{V}_X^T\hat{\tilde{\Z}}\\\nonumber  
    (d)\quad&\hat{\L},\hat{\tilde{\Z}} \in \argmin_{\substack{\tilde{\Z}\in\mathcal{S}(P,K),\\ \L\in\mathcal{M}(P,K)}}
    \left(\frac{1}{2}\|\C-\tilde{\Z}\L^T\|_F^2
    + \sum_{k=1}^K P(\l_k; \lambda_k)\right)
    \\\nonumber&\text{and set }\hat{\Z}=\mathbf{U}_X\mathbf{V}_X^T\hat{\tilde{\Z}}.
\end{align}
where $\sum_{k=1}^K P(\l_k; \lambda_k)$ is an arbitrary penalty term on $\l_k$ with parameter $\lambda_k$.
\end{theorem}

\begin{proof}
The equivalence of (a) and (b) follows from Theorem \ref{ebcd_thm:equivalence_main}. (b) and (c) are equivalent because $\mathbf{X}^T\mathbf{X}=\C^T\C$ and $\text{Polar.U}(\mathbf{Q}\mathbf{UDV}^T)=\mathbf{QUV}^T=\mathbf{Q}\text{Polar.U}(\mathbf{UDV}^T)$ for any $\mathbf{Q}$ that satisfies $\mathbf{Q}^T\mathbf{Q}=\mathbf{I}$. And the equivalence of (c) and (d) again follows from Theorem \ref{ebcd_thm:equivalence_main}. 
\end{proof}

From Theorem \ref{ebcd_thm:reformulation}, any penalized PCA criterion $(a)$ can be reformulated in the form $(d)$, in which the target matrix $\mathbf{X}\in\mathcal{M}(N,P)$ is replaced by a compact version $\C\in\mathcal{M}(P,P)$. This can then be solved by applying the BISPCA algorithm to $\C$. If the component score matrix $\Z$ is not a parameter of interest then no additional step is needed; otherwise, $\Z$ can be easily recovered using the singular vectors of $\mathbf{X}$. If $P\ll N$ then this approach may be computationally more efficient than directly applying the BISPCA algorithm to $\mathbf{X}$.

The potential to use the Gram matrix for efficient computation, has been previously stated in the context of specific sPCA models, for example in \cite{Journee.Nesterov.ea2010}.
 Our contribution is to provide a general result that applies to any penalty function, which enables applications to not-so-straightforward problems (e.g. EBCD in Section \ref{ebcd_sec:ebcd}).

\section{Proof of Proposition \ref{ebcd_prop:algo}}\label{ebcd_sec:algoproof}

\begin{proof}
The evidence lower bound (ELBO) of the model, $F(\g, \Z, \tau, \q)$, can be written as
\begin{align}\label{ebcd_eq:elbo_full}
    F(\g, \Z, \tau, \q) =& -\frac{NP}{2}\log(2\pi) + \frac{NP}{2}\log(\tau) \\\nonumber &- \frac{\tau}{2}\mathbb{E}_{\mathbf{q}} \left[ \|\X - \Z\L^T\|_F^2 \right] + \mathbb{E}_{\mathbf{q}}\left[\log\frac{\mathbf{g}(\L)}{\mathbf{q}(\L)}\right],
\end{align}
and the three steps iteratively maximizing the ELBO can be shown as follows. (a) \textit{EBNM step}: maximizing ELBO with respect to $(\mathbf{g}, \mathbf{q})$ factorizes into $K$ subproblems of the form $\max_{(g_k, q_k)}\mathbb{E}_{q_k}\left[\log\frac{g_k(\l_k) \prod_p \exp(-\frac{\tau}{2}(l_{p,k}-(\X^T\Z)_{p,k})^2)}{q_k(\l_k)}\right]$, which corresponds to the EBNM problem
$\text{EBNM}(\mathbf{X}^T\zk, 1/{\tau}, \mathcal{G})$.
(b) \textit{Rotation step}: maximizing ELBO with respect to $\Z$ reduces to a reduced-rank Procrustes rotation problem, $\min_\Z \|\X-\Z\bar{\L}^T\|_F^2$, which has the solution $\text{Polar.U}(\X\bar{\L})$.
(c) \textit{Precision step}: maximizing ELBO with respect to $\tau$ has the closed form solution $\tau=NP/\mathbb{E}_{\mathbf{q}} \left[ \|\X - \Z\L^T\|_F^2 \right]=NP/(\|\mathbf{X}-{\Z}{\bar{\L}} \text{}^T\|_F^2+\|{\mathbf{V}}\|_{1,1})$.
\end{proof}

\section{Proof of Proposition \ref{ebcd_prop:EB_cov_interpretation}}
\begin{proof}
The evidence lower bound (ELBO) of the model, $F(\g, \Z, \tau, \q)$ in \eqref{ebcd_eq:elbo_full}, can be rearranged as 
\begin{align}
    &-\frac{\tau}{2}\|\mathbf{X}-\mathbf{Z}\bar{\mathbf{L}}\|_F^2 \\\nonumber &- \tau \sum_{p,k}\left(
    \frac{N}{2K\tau}\log\left(\frac{2\pi}{\tau}\right)
     + \frac{1}{2}\left( var_{q_{p,k}}(l_{p,k}) + \frac{2}{\tau}\mathbb{KL}(q_{p,k}||g_k)    \right)
    \right),
\end{align}
and after taking the maximum over $\mathbf{q}: \mathbb{E}[\mathbf{L}]=\bar{\mathbf{L}}$, we get the expression \eqref{ebcd_eq:P_tau}.
\end{proof}

\section{Extensions and variations} \label{ebcd_sec:ext}

\subsection{Scaled versions of the sparse PCA criterion}
\label{ebcd_sec:scaling}

One slightly unnatural feature of the formulations presented in the main text
is that they place a penalty (or prior) on a parameter, $\L$, that is not a ``population quantity", and whose interpretation changes with the number of samples $N$. For example, in Section \ref{ebcd_sec:ebcd} we saw that the fidelity term encourages $\L\L^T \approx \X^T\X$, whose magnitude grows with $N$; 
it would seem more natural to combine a penalty on $\L$ with a fidelity term that encourages $\L\L^T \approx (1/N)\X^T\X$ since the latter has a natural limit as $N \rightarrow \infty$ (with $P$ fixed).
This can be achieved simply by replacing the constraint $\Z^T\Z=\mathbf{I}_K$ with the scaled version $\Z^T\Z=N\mathbf{I}_K$, or equivalently $\Z/\sqrt{N} \in \mathcal{S}(N,K)$. All our results and algorithms are easily modified for this rescaled version. For example, the sparse PCA criterion \eqref{ebcd_eq:penpca} becomes
\begin{equation} 
    \min_{\substack{\Z/\sqrt{N} \in\mathcal{S}(N,K),\\ \L\in\mathcal{M}(P,K)}}
    \left(\frac{1}{2}\|\mathbf{X}-\Z\L^T\|_F^2 +
    \sum_{k=1}^K P(\l_k; \lambda_k) \right);
\end{equation}
the equivalent covariance formulation (\eqref{ebcd_eq:pencov_main}
and (b) in Theorem \ref{ebcd_thm:reformulation}) becomes
\begin{equation}  \label{ebcd_eq:pencov_scaled}
   \min_{\L\in\mathcal{M}(P,K)} \left(\frac{N}{2}d_*(\mathbf{X}^T\mathbf{X}/N, \L\L^T)^2+\sum_{k=1}^K P(\l_k; \lambda_k) \right);
\end{equation}
the equivalent compact matrix formulation ((d) in Theorem \ref{ebcd_thm:reformulation}) becomes
\begin{equation}
\argmin_{\substack{\tilde{\Z}/\sqrt{N} \in \mathcal{S}(P,K),\\ \L\in\mathcal{M}(P,K)}}
    \left(\frac{1}{2}\|\C-\tilde{\Z}\L^T\|_F^2
    + \sum_{k=1}^K P(\l_k; \lambda_k)\right);
\end{equation}
and the penalty term \eqref{ebcd_eq:P_tau} becomes
\begin{align} \label{ebcd_eq:pen_revised}
    P_{\tau,g}(\bar{l})
    =&
    \frac{N}{2K\tau}\log\frac{2\pi}{\tau}\\\nonumber
    &+ \frac{1}{2}\min_{
                q: \mathbb{E}_{q}[l]=\bar{l}
    } \left(N var_{q}(l) + \frac{2}{\tau} \KL(q||g) \right).
\end{align}
The BISPCA updates become
\begin{equation}
\l_k\gets S_{\rho/N}(\mathbf{X}^T\zk/N; \lambda_k); \qquad
\Z \gets \sqrt{N}\text{Polar.U}(\mathbf{X}\L);
\end{equation}
and the EBCD updates \eqref{ebcd_eq:ebnm_step}-\eqref{ebcd_eq:precision_step} become
\begin{align} 
&\text{EBNM step: }\text{for each }k\in [K], \\\nonumber 
&\quad\quad\quad(g_k,q_k) \gets \text{EBNM}(\mathbf{X}^T\zk/N, 1/{N\tau}, \mathcal{G})\\
&\text{Rotation step: } 
{\Z} \gets \sqrt{N}\text{Polar.U}(\mathbf{X}{\bar{\L}})\\ 
&\text{Precision step: }
{\tau} \gets 
NP/(\|\mathbf{X}-{\Z}{\bar{\L}} \text{}^T\|_F^2+N\|{\mathbf{V}}\|_{1,1}) \\\nonumber
&\quad\quad\quad\quad\quad\quad\quad\quad \left[= P/(d_*(\X^T\X/N,\L \L^T)^2 + \|{\mathbf{V}}\|_{1,1}) \right]. 
\end{align}
And, just as before, one can apply these updates to a compact version of the data matrix to solve the same problem. 

This modification to the methods makes it easier to reason about their behavior in the regime $N \rightarrow \infty$ with $P$ fixed, where we can assume $\lim_{N \rightarrow \infty} \X^T\X/N = \mathbf{S}$ say.  For example, \eqref{ebcd_eq:pencov_scaled} shows that for a fixed penalty (not depending on $N$) the influence of the penalty will decrease as $N$ increases, and the limiting estimate of $\L$ will be $\in \argmin d_*(\mathbf{S}, \L \L^T)$ independent of the penalty. And because the part of the penalty \eqref{ebcd_eq:pen_revised} depending on $g$ does not scale with $N$, the effect of the prior $g$ diminishes as $N \rightarrow \infty$ as one might expect (indeed, in the limit as $N \rightarrow \infty$ the EBCD optimum $\L$ will be $\in \argmin d_*(\mathbf{S}, \L \L^T)$ whether $g$ is fixed or estimated from the data).

\subsection{Column-wise variances} 
\label{ebcd_sec:hetero}

We can extend the EBCD model \eqref{ebcd_eq:model_start}-\eqref{ebcd_eq:model_end} to allow different variables to have different variances/precisions:
\begin{align}
    \mathbf{X}&=\Z\L^T+\mathbf{E}\\ 
    l_{p,k} &\sim^\text{iid} g_k\in\mathcal{G}\\ 
    e_{n,p} &\sim^\text{iid} N(\cdot; 0, 1/{\tau_p})
\end{align}
where $\Z\in\mathcal{S}(N,K)$. Fitting this heteroskedastic model requires solutions for the heteroskedastic versions of the reduced-rank Procrustes rotation problem and the EBNM problem, as we now detail.

\begin{fact}[Heteroskedastic Reduced-rank Procrustes rotation problem] \label{ebcd_fact:proc_het}
Given $\L$, the minimum
$$\min_{\Z\in\mathcal{S}(N,K)} \sum_{n,p} \tau_p (x_{n,p}-(\Z\L^T)_{n,p})^2
$$
is achieved by $\Zhat(\L,\X,\T):=\text{Polar.U}(\mathbf{X}\T\L)$
where $\T$ is the $P \times P$ diagonal matrix with $T_{p,p}=\tau_p$.
\end{fact}
\begin{proof}
The minimization problem is equivalent to $\min_{\Z\in\mathcal{S}(N,K)}\|(\X-\Z\L^T)\sqrt{\mathbf{T}}\|_F^2= \min_{\Z\in\mathcal{S}(N,K)}$ 
$\|\X\sqrt{\mathbf{T}}-\Z(\sqrt{\mathbf{T}}\L)^T\|_F^2$, which reduces to a (homoskedastic) reduced-rank Procrustes rotation problem and has a solution $\text{Polar.U}(\X\sqrt{\mathbf{T}}\sqrt{\mathbf{T}}\L)=\text{Polar.U}(\X\mathbf{T}\L)$, where $\sqrt{\mathbf{T}}$ is the $P \times P$ diagonal matrix with diagonal entries $\sqrt{\tau_p}$.
\end{proof}

\begin{definition} \label{ebcd_def:ebnm_het}
Let $\text{EBNM}(\x,\s^2,\G)$ denote a function that returns the EB solution to the following heteroskedastic normal means model:
    \begin{align} 
    x_p|\eta_p, s_p^2 &\sim^\text{indep} N(x_p; \eta_p,s_p^2) \\
    \eta_p &\sim^\text{iid}g \in \G,
\end{align}
for $p=1,\dots,P$.  
\end{definition}

\begin{proposition}\label{ebcd_prop:algo_het}
Maximizing the evidence lower bound $F(\g, \Z, \T, \q)$ (\eqref{ebcd_eq:elbo} but with $\tau$ replaced by $\T$) subject to $\Z^T\Z=\mathbf{I}_K$ can be achieved by iteratively updating $(\g,\mathbf{q})$, updating $\Z$, and updating $\T$, as follows:
\begin{align} 
&\text{EBNM step: }\text{for each }k\in [K], \\\nonumber 
&\quad\quad\quad (g_k,q_k) \gets \text{EBNM}(\mathbf{X}^T\zk, (1/\tau_1,\dots,1/\tau_P), \mathcal{G})\\
&\text{Rotation step: } 
{\Z} \gets \text{Polar.U}(\mathbf{X}\T{\bar{\L}})\\
&\text{Precision step: }\text{for each }p \in [P], \\\nonumber
&\quad\quad\quad \tau_p \gets 
N/ \left(\sum_n (x_{n,p} -({\Z}{\bar{\L}})_{n,p})^2 +\sum_k v_{p,k} \right).
\end{align}
Here $\bar{\L} = \E_\q(\L)$ and $v_{p,k}=\text{Var}_{q_k}(l_{p,k})$.
\end{proposition}
Note that in practice, one would need to apply some regularization when estimating $\tau_p$ to prevent solutions with $\tau_p \rightarrow \infty$.

\end{document}